\documentclass[aps,prl,amsmath,amssymb,twocolumn,showpacs,showkeys,superscriptaddress, floatfix]{revtex4-1}

\usepackage{graphicx}
\usepackage{dcolumn}
\usepackage{color}
\usepackage{ulem}
\usepackage{float}
\usepackage{hyperref}

\newcommand{\ekp}{eight-band \mbox{${\bf k}\!\cdot\!{\bf p}$}}
\begin{document}
\title{Electronic states of (InGa)(AsSb)/GaAs/GaP quantum dots}

\author{Petr Klenovsk\'y}
\email[]{klenovsky@physics.muni.cz}
\affiliation{Department of Condensed Matter Physics, Faculty of Science, Masaryk University, Kotl\'a\v{r}sk\'a~267/2, 61137~Brno, Czech~Republic}
\affiliation{Central European Institute of Technology, Masaryk University, Kamenice 753/5, 62500~Brno, Czech~Republic}
\affiliation{Czech Metrology Institute, Okru\v{z}n\'i 31, 63800~Brno, Czech~Republic}




\author{Andrei Schliwa}
\affiliation{Institute for Solid State Physics, TU Berlin,  Germany}

\author{Dieter Bimberg}
\affiliation{Institute for Solid State Physics, TU Berlin,  Germany}
\affiliation{Bimberg Chinese German Center for Green Photonics of the Chinese Academy of Sciences at CIOMP, Changchun, Jilin, Peoples R China}

\date{\today}

\begin{abstract}

 Detailed theoretical studies of the electronic structure of (InGa)(AsSb)/GaAs/GaP quantum dots are presented. This system is unique since it exhibits concurrently direct and indirect transitions both in real and momentum space and is attractive for applications in quantum information technology, showing advantages as compared to the widely studied (In,Ga)As/GaAs dots. We proceed from the inspection of the confinement potentials for ${\bf k}\neq 0$ and ${\bf k}= 0$ conduction and ${\bf k}= 0$ valence bands, through the formulation of ${\bf k}\cdot{\bf p}$ calculations for ${\bf k}$-indirect transitions, up to the excitonic structure of $\Gamma$-transitions. Throughout this process we compare the results obtained for dots on both GaP and GaAs substrates, enabling us to make a direct comparison to the (In,Ga)As/GaAs quantum dot system. We also discuss the realization of quantum gates.

\end{abstract}

%

\maketitle

\section{Introduction}

Monolithic integration of III-V compounds with Si-technology is one of the key challenges of future photonics~\cite{liang_recent_2010}. The problems caused by the large lattice mismatch between Si and typical emitter materials based on GaAs or InP substrates can be avoided to a large extent by employing a pseudomorphic approach, i.e. growing almost lattice matched compounds on Si. The III-V binary compound with the lattice constant closest to Si is GaP (0.37\% lattice mismatch at 300K). GaP is an indirect semiconductor, and thus not seen as a useful laser  material. It might serve however as a matrix for more appropriate material combinations. The initially obvious choice of employing InGaP as active material fails due to the borderline type-I/II nature of the bandoffset to GaP~\cite{hatami_inp_2003}. (In,Ga)As/GaP, by contrast, features a type-I lineup and triggered a fair amount of research, both experimental and theoretical in nature~\cite{leon_self-forming_1998,guo_first_2009,shamirzaev_high_2010,umeno_formation_2010,fuchi_composition_2004,nguyen_thanh_room_2011-1,rivoire_photoluminescence_2012}. The main issue with this material combination is the large lattice mismatch and the resulting large strain in  the (In,Ga)As active material, possibly leading to direct - indirect crossover of the ground state transition. Fukami et al.~\cite{fukami_analysis_2011} were the first to evaluate the necessary fraction of In for a direct electron-hole ground state transition using model-solid theory for (InGa)(AsN)/GaP. Further theoretical insight was provided by the work of Robert et al.~\cite{Robert2012,Robert2014,Robert2016} who first employed a mixed ${\bf k}\cdot{\bf p}$ / tight-binding simulations, predicting a direct-indirect crossover at about 30\% In-content for larger (In,Ga)As/GaP quantum dots (QDs). For smaller QDs they predicted an even larger In content for the direct transition in reciprocal space.

\begin{figure}[!ht]
	\begin{center}
		\begin{tabular}{c}
			\includegraphics[width=0.45\textwidth]{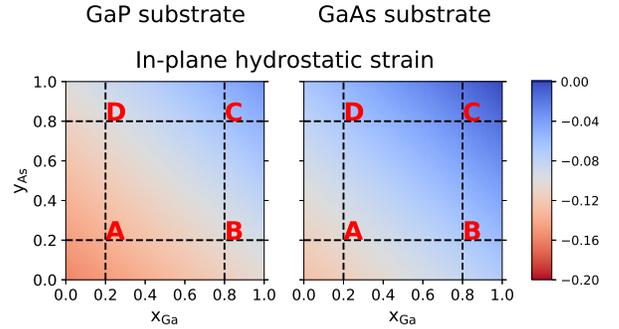}
		\end{tabular}
	\end{center}
	\caption{In-plane hydrostatic strain of In$_{1-x}$Ga$_x$As$_y$Sb$_{1-y}$ alloy, lattice matched to GaP (GaAs) is shown in the left (right) panel. The capital letters A, B, C, and D mark the concentrations listed in Tab.~\ref{tab:ABCDdesig}. We also introduce here the notation $x_{\rm{Ga}}$ and $y_{\rm{As}}$ marking the content of Ga and As, respectively, in In$_{1-x}$Ga$_x$As$_y$Sb$_{1-y}$ alloy. For the interpolation scheme between different constituents used here see Eqs.~(\ref{Eq:alloy_lin-1})~and~(\ref{Eq:alloy_tern-1}). Notice the pronounced compressive stress towards pure InSb for structures grown on GaP.
		%
		\label{fig:GaPVSGaAsSubs}}
\end{figure}

In the present work we take the next step and assess the role of additional antimony incorporation, leading to In$_{1-x}$Ga$_x$As$_y$Sb$_{1-y}$/GaP QDs based on the experimental works of Sala et al.~\cite{Sala2016,t_sala,Sala2018}. Not only will we look at its suitability as optoelectronic material~\cite{ArxivSteindl:19} but also - as discussed  by Sala et al.~\cite{t_sala,Sala2018} - as material for QD-Flash memories.

The QD-Flash memory concept was suggested and developed by Bimberg {\sl et al.} over a period of 20 years following the first studies by Kapteyn {\sl at al.} on electron escape mechanism from InAs QDs using the deep level transient spectroscopy (DLTS)~\cite{Kapteyn1999,Marent2011,Bonato2016,BimbergPatent}. The concept, protected by 16 patents worldwide, attempts to combine the best of both memory worlds, the Dynamic Random Access Memories (DRAM) and the Flash worlds leading to a universal memory, strongly simplifying computer architecture. Fast read-write-erase operations, as fast or faster than those in current DRAM, shall be combined with non-volatility of information for more than 10 years in the same device. Presently most promising storage elements are of type-II QDs storing solely holes. GaSb QDs embedded in GaP show hole retention times of 4 days and the limit of 10 years is predicted to be crossed by varying the structures to (In,Ga)Sb QDs embedded in (Al,Ga)P.
%
%

The secret for successful growth of such In$_{1-x}$Ga$_x$As$_y$Sb$_{1-y}$/GaP QDs by MOCVD constitutes a 5 ML GaAs interlayer (IL) on top of the GaP matrix material, thus, enabling QD formation~\cite{Sala2016,t_sala}, which will be carefully considered in the following simulations. The choice of GaAs layer is evident from Fig.~\ref{fig:GaPVSGaAsSubs} where we compare the effects of the GaP and more conventional GaAs substrates on hydrostatic strain in hypothetical bulk lattice-matched In$_{1-x}$Ga$_x$As$_y$Sb$_{1-y}$ alloy. Note that Fig.~\ref{fig:GaPVSGaAsSubs} highlights also the labelling convention used in this work in order to avoid confusion: the Ga content in In$_{1-x}$Ga$_x$As$_y$Sb$_{1-y}$ is marked as $x_{\rm{Ga}}$ while that for As is $y_{\rm{As}}$.



\section{General remarks and outline}

In our system, compared to~e.g.~(In,Ga)As/GaAs QDs, the ${\bf k}$-indirect electron states attain lower energy than the $\Gamma$ ones. This is a result of the large compressive strain in QDs occurring due to GaP substrate. Moreover, the eight- (six-) fold symmetry of L (X) bulk Bloch waves translates into four- (three-) L (X) envelope functions for quasiparticles in QDs, 
since each state is shared by two neighbouring Brillouin zones. We denote the resulting envelope wavefunctions L$_{[110]}$, L$_{[\bar{1}\bar{1}0]}$, L$_{[1\bar{1}0]}$, and L$_{[\bar{1}10]}$ (X$_{[100]}$, X$_{[010]}$, and X$_{[001]}$). The degeneracy of envelopes for L$_{[110]}$, L$_{[\bar{1}\bar{1}0]}$, L$_{[1\bar{1}0]}$, and L$_{[\bar{1}10]}$, or X$_{[100]}$, X$_{[010]}$, and X$_{[001]}$ bands is lifted in real dots due to structural imperfections (e.g. shape, composition) or by external perturbations (e.g. electric, magnetic, or strain fields) and we thus distinguish between these bands in the following, and study also the effects of degeneracy lifting.
We carefully choose three exemplary points A, B, C, and D as seen in Fig.~\ref{fig:GaPVSGaAsSubs} and Tab.~\ref{tab:ABCDdesig}, that exhibit certain specific properties of our system, which will be discussed further in the the body of the paper.
\begin{table}[!ht]
	\centering
	\caption{Ga and As concentrations corresponding to points A, B, C, and D in Fig.~\ref{fig:GaPVSGaAsSubs}.}
	\label{tab:ABCDdesig}
	\begin{tabular}{c|c}
		{\bf A}  &  $x_{\rm{Ga}}$ = 0.2; $y_{\rm{As}}$ = 0.2\\
		{\bf B}  &  $x_{\rm{Ga}}$ = 0.8; $y_{\rm{As}}$ = 0.2\\
		{\bf C}  &  $x_{\rm{Ga}}$ = 0.8; $y_{\rm{As}}$ = 0.8\\
		{\bf D}  &  $x_{\rm{Ga}}$ = 0.2; $y_{\rm{As}}$ = 0.8\\
	\end{tabular}
\end{table}

\begin{figure}[!ht]
	\begin{center}
		\begin{tabular}{c}
			\includegraphics[width=0.48\textwidth]{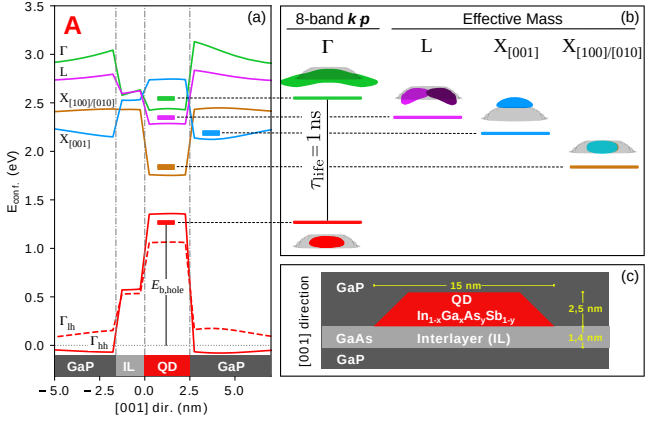} 
		\end{tabular}
	\end{center}
	\caption{Schematic overview of the presented results: (a) bandedges of $\Gamma$, L, X$_{[001]}$, X$_{[100]/[010]}$ electron and $\Gamma$ hole bands for QD with $x_{\rm{Ga}}$ = 0.2 and $y_{\rm{As}}$ = 0.2, marked as A in Tab.~\ref{tab:ABCDdesig}. The corresponding single-particle ground state eigen energies are indicated by thick horizontal lines and correspond in panel (b) to side views of the probability densities of the envelope functions. QD body in (b) is indicated by grey objects. The top panel in (b) shows the method of calculation of ${\bf k}=0$ and ${\bf k}\neq 0$ states in our theory. The vertical line between $\Gamma$ electron and hole states marks the recombination between these states with radiative lifetime of $\tau_{\rm life}=1\,\,{\rm ns}$. In panel (c) we give the side view of the simulated In$_{1-x}$Ga$_x$As$_y$Sb$_{1-y}$/GaAs/GaP QD. The shape of QD is a truncated cone with height $h=2.5$~nm and base (top) diameter $d_b=15$~nm ($d_t=8$~nm). The QDs are positioned on a 5~ML thick IL of pure GaAs and the whole structure is embedded in GaP.
		\label{fig:QDsketch}}
\end{figure}

The paper is organized as follows: first we introduce our method of calculation. Single-particle states are calculated as a combination  of one-band (for L- and X-point states) and eight-band $\mathrm{{\bf k}}\cdot{ \mathrm{\bf p}}$ approximation (for $\Gamma-$point states) \{see top inset of Fig.~\ref{fig:QDsketch}~(b) and Fig.~\ref{fig:MOC}\}. Owing to the very large lattice-mismatch between GaP and the QD-material, a method for the calculation of the inhomogeneous strain and its impact on the local bandedges is introduced, together with the effect of piezoelectricity. Our methods for accounting Coulomb interaction and calculation of optical properties are introduced thereafter.

Next, we continue with the analysis of the arising confinement potentials \{Fig.~\ref{fig:QDsketch}~(a)\} and analyze the electron and hole probability densities and eigen energies, respectively \{Fig.~\ref{fig:QDsketch}~(b)\}. Based on these results, we then inspect the electron-hole Coulomb integrals for $\Gamma$-point states and derive information on type-I/II behavior. Then we discuss the localization energies of holes in our dots, which are relevant for the QD-flash memory concepts. We continue by studying the emission properties and the fine-structure of those excitons consisting of $\Gamma$-electrons and holes. Finally, we present an application of In$_{1-x}$Ga$_x$As$_y$Sb$_{1-y}$/GaAs/GaP QD system as a possible realization of quantum gate and briefly discuss its  properties.

\section{Method of calculation}

\begin{figure}[!ht]
	\begin{center}
		\begin{tabular}{c}
			\includegraphics[width=0.3\textwidth]{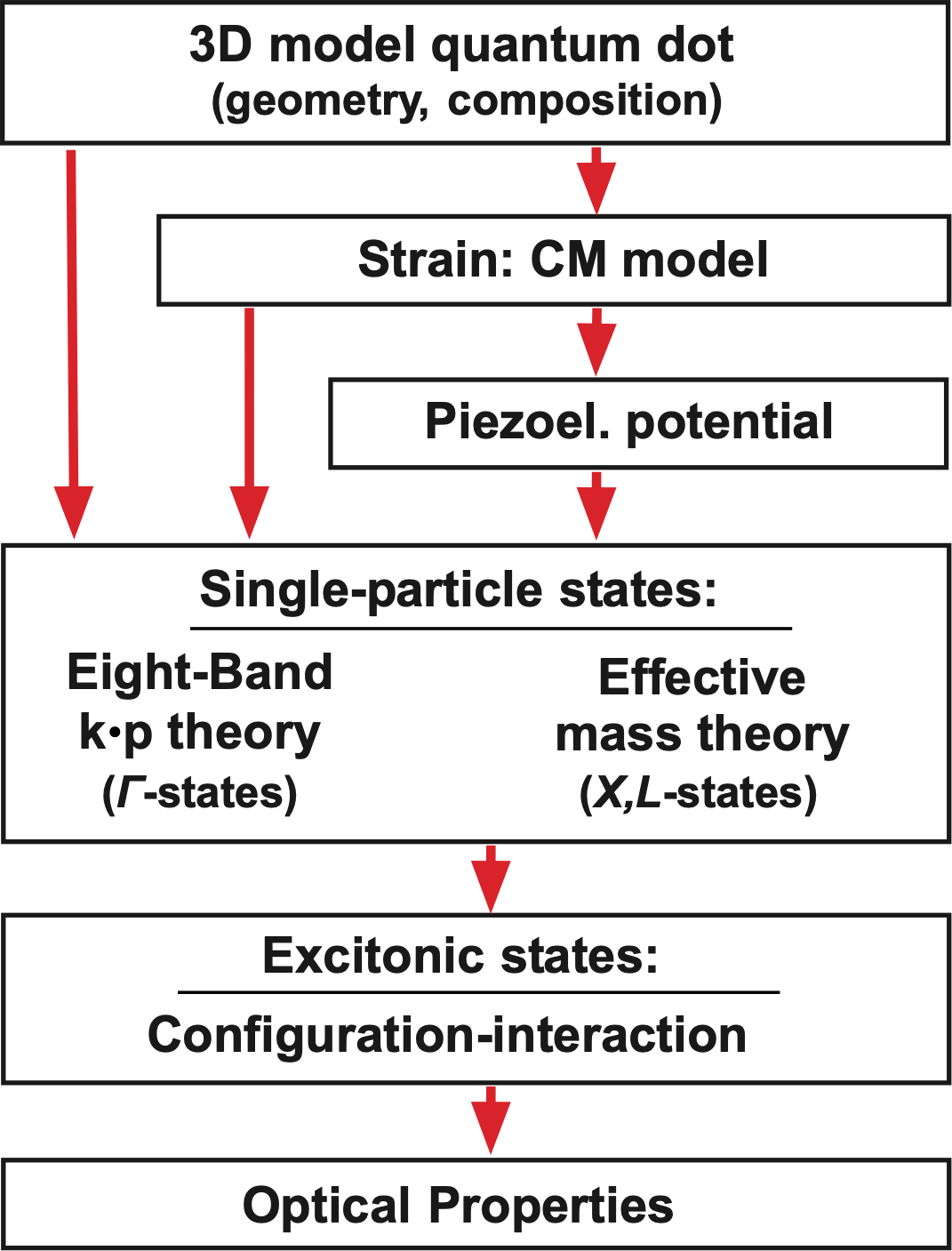}
		\end{tabular}
	\end{center}
	\caption{Schematics of the modeling procedure applied in this work.
		\label{fig:MOC}}
\end{figure}
Figure~\ref{fig:MOC} shows an outline of the modelling procedure employed in this paper. It starts with an implementation of the 3D QD model structure (size, shape, chemical composition), and carries on  with the calculation of strain and piezoelectricity. The resulting strain and polarization fields then either enter the eight-band $\mathbf{k}\!\cdot\!\mathbf{p}$ Hamiltonian for states located around the Brillouin-zone-center ($\Gamma$-point), or the effective-mass Hamiltonian for those emerging off-center such as L- and X-point states. Solution of the resulting Schr\"odinger equations yields electron and hole single-particle states both at the $\Gamma$- as well at X- and L-points. Coulomb interaction is accounted for by employing the configuration interaction (CI) method including dipole-dipole interaction. Finally, optical properties such as absorption spectra, capture cross sections, or lifetimes can be calculated.

\subsection{Choice of model structure}

The morphology of our model QD is related to the works of Stracke and Sala \cite{Stracke2014,Sala2016,t_sala,Sala2018}: The whole structure is grown on GaP substrate with an IL between QD and substrate made of 5 ML GaAs \{see Fig.~\ref{fig:QDsketch}~(c)\}. Generally, the IL
is of critical importance to enable the QD formation, as discovered by Stracke and coworkers for the In$_{1-x}$Ga$_{x}$As/GaAs/GaP QDs~\cite{stracke_growth_2012,Stracke2014}. There, the IL thickness used was around 2-3 ML, which remarkably affected the GaP surface reconstruction and diffusion, eventually enabling the QD formation. Similarly, for In$_{1-x}$Ga$_{x}$As$_{y}$Sb$_{1-y}$/GaAs/GaP QDs, the GaAs IL is used to enable the QD growth, but its thickness is of about 5ML~\cite{Sala2016,t_sala}. Here, it's very likely that an intermixing via As-for-Sb exchange between the GaAs IL and the Sb of the QDs takes place, such that part of the IL becomes part of the QDs. Such process may lower the strain between QDs and GaP, where nominally the lattice mismatch was very high (nominally of $\sim$ 13 $\%$ between In$_{0.5}$Ga$_{0.5}$Sb/GaP) for enabling an usual Stranski-Krastanov QD growth. Therefore, such intermixing may have lowered the high mismatch, thus enabling the QD formation \cite{Sala2016,t_sala,Sala2018}, similarly observed also in Abramkin \textit{et al.}~\cite{Abramkin2014} for GaSb/GaP QDs. Note, that we do not consider the described intermixing of Sb to the IL in order to make our results more general, and not depending on particular QD growth conditions.


The square based QD itself is made of In$_{1-x}$Ga$_{x}$As$_{y}$Sb$_{1-y}$ with a base-length of 15 nm and a height of 2.5 nm, based on realistic QD features~\cite{Sala2016,t_sala,Sala2018}. We use constant atomic distribution of the constituents of the QD in our work. While it is known that an alloy gradient is important for the built-in electron-hole dipole moment~\cite{Fry:00,Grundmann:95,Klenovsky2018} it has a rather small impact on emission energy or fine-structure of exciton~\cite{Klenovsky2018}, which will be discussed in the following. 

\subsubsection{Alloying}

To properly describe the In$_{1-x}$Ga$_{x}$As$_{y}$Sb$_{1-y}$
alloy, we used in all steps of the aforementioned procedure the following
interpolation equation~\cite{Birner:07} 
\onecolumngrid

\begin{subequations} 
	\begin{align}
		f_{\mathrm{quat}}(x,y)= & (1-x)yf_{\mathrm{InAs}}+xyf_{\mathrm{GaAs}}+(1-x)(1-y)f_{\mathrm{InSb}}+x(1-y)f_{\mathrm{GaSb}}\label{Eq:alloy_lin-1}\\
		& +x(1-x)yf_{\mathrm{InGa,As}}+x(1-x)(1-y)f_{\mathrm{InGa,Sb}}+(1-x)y(1-y)f_{\mathrm{AsSb,In}}+xy(1-y)f_{\mathrm{AsSb,Ga}}\label{Eq:alloy_tern-1}
	\end{align}
\end{subequations} 
\twocolumngrid
where Eq.~(\ref{Eq:alloy_lin-1})
gives the linear and Eq.~(\ref{Eq:alloy_tern-1}) the quadratic material
interpolation parameters, respectively. 
For the full list of material parameters used in this work see Ref.~\cite{SupMatPRBgen}~(see, also, references~\cite{LUTTINGER1956,Varshni1967,DRESSELHAUS1955,Wei1998,Wei1999,WALLE1989} therein).

\subsection{Single-particle states}

Owing to the choice of materials and the arising large strain values,
the conduction band electronic ground state is in general not a $\Gamma$-state.
Hence, we resort to a hybrid approach~\cite{KlenovskyPRB2012} where we
calculate the $\Gamma$-states using the eight-band $\mathbf{k}\!\cdot\!\mathbf{p}$-model, and the L- and X-states using the effective mass model, both including strain and piezoelectricity. All the preceding steps of the calculation are done using the nextnano$++$ simulation suite~\cite{Birner:07,t_zibold}.

%
%
The choice of different models here is motivated by the relative smallness of the coupling parameter between ${\bf k}\neq 0$ conduction and ${\bf k}= 0$ valence Bloch states, respectively, which allows us to approximately decouple transitions involving ${\bf k}\neq 0$ conduction band (CB) from $\Gamma$-valence bands (VBs) and, thus, treat the former by effective mass approach. The general reason is the emission probability $R^E_{\mathrm{ind},{\bf k}}$ of such an event in bulk indirect semiconductor in the low temperature limit ($N_{\rm p}+1\approx 1$ where $N_{\rm p}$ is the Bose-Einstein statistics) reads
%
\begin{equation}
\label{Eq:KaneParamIndirect}
R^E_{\mathrm{ind},{\bf k}}\sim\sum_j\left|\sum_i\frac{\left<u_v^{\Gamma}\left|\mathcal{H}_{\mathrm{eR}}\right|i\right>\left<i\left|\mathcal{H}_{\mathrm{ep}}\right|u_c^{{\bf k}}\right>}{E_{i\Gamma}-E_{\rm ind}-\hbar\omega_j({\bf k})}\right|^2\,.
%
%
\end{equation}
%
where $i$ and $j$ label the virtual states and the phonon branches for ${\bf k}$, respectively, $u_v^{\Gamma}$ and $u_c^{\bf k}$ mark Bloch waves in ${\bf k}=0$ of valence and ${\bf k}\neq 0$ of conduction band, respectively, 
$\mathcal{H}_{\mathrm{ep}}$ and $\mathcal{H}_{\mathrm{eR}}$ are Hamiltonians for the electron-phonon and electron-photon interaction, respectively, $E_{i\Gamma}$ is the energy of the $i$-th virtual state at $\Gamma$-point, $E_{\rm ind}$ is the bandgap of the indirect semiconductor, and $\omega_{j}(\bf k)$ marks the frequency of $j$-th phonon branch corresponding to momentum $\bf k$; $\hbar$ marks the reduced Planck's constant.
Equation~(\ref{Eq:KaneParamIndirect}) is derived in Ref.~\cite{SupMatPRBgen}~(see, also, references~\cite{LUTTINGER1956,Varshni1967,DRESSELHAUS1955,Wei1998,Wei1999,WALLE1989} therein) and it is based on equation (6.61) in Ref.~\cite{YuCardona} describing the light absorption in indirect semiconductors, the general theory is on the other hand worked out,~e.~g.,~in~\cite{Dirac:58, Landau:65}.
For comparison, in similar fashion the probability for transition in $\Gamma$-point of direct semiconductor (Fermi's Golden Rule) reads
\begin{equation}
\label{Eq:KaneParamDirect}
R^E_{\mathrm{dir},\Gamma}\sim\left|\left<u_v^{\Gamma}\left|\mathcal{H}_{\mathrm{eR}}\right|u_c^{\Gamma}\right>\right|^2\,.
\end{equation}
%

The elements of the kind of Eqs.~(\ref{Eq:KaneParamIndirect})~and~(\ref{Eq:KaneParamDirect}) are usually obtained by atomistic theories like the Density Functional Theory, empirical pseudopotentials~\cite{Wang1997,Wang1997a,Robert2016}, or others and their evaluation is not the scope of this work. However, clearly the probability given by Eq.~(\ref{Eq:KaneParamIndirect}) is expected to be much smaller than for Eq.~(\ref{Eq:KaneParamDirect}) owing to the necessity of the involvement of virtual states and coupling to phonons in the former case. Since Eq.~(\ref{Eq:KaneParamDirect}) is the basis for computing the Kane's parameter $P_{\Gamma}$ employed in eight-band ${\bf k}\cdot{\bf p}$ method to describe the coupling of CB and VBs at $\Gamma$ point, it is reasonable to assume that a similar element for ${\bf k}$-indirect transition $P_{{\bf k}\neq 0}$ based on Eq.~(\ref{Eq:KaneParamIndirect}) will be much smaller than $P_{\Gamma}$, finally leading to our choice of the methods of ${\bf k}\cdot{\bf p}$ calculation for direct and indirect states and we, thus, also set $P_{{\bf k}\neq 0}=0$. We note that our choice is verified by the results of Refs.~\cite{Landsberg1973}~and~\cite{VarshniRate1967} using which we estimate the upper limit $P_{{\bf k}\neq 0}/P_{\Gamma}<10^{-3}$ for all bulk semiconductor constituents of In$_{1-x}$Ga$_{x}$As$_{y}$Sb$_{1-y}$ alloy.

Another possible issue arises from mixing between individual CB states, like between L$-\Gamma$, X$-\Gamma$, and L-X. Here we resort to the results of Ref.~\cite{Robert2016}~and~\cite{Wang1997} where that is computed for similar QD structures. In fact, the magnitude of mixing discussed in those works seems to be rather small,~i.~e.,~$\sim 10^{-2}$~\%~\cite{Robert2016}. Furthermore, Wang~{\sl et al.}~\cite{Wang1997a} found that the mixing is smaller in QDs than in higher dimensional structures. Moreover, the conclusion that our QDs are too large for CB mixing to be of considerable importance, can be deduced from results of Ref.~\cite{Diaz2006}. Hence, since we aim in this work on general properties of the studied system, it is reasonable to omit mixing between CB states here even though we note that a fuller description should be obtained when that is taken into account. 

%

\subsubsection{Eight band $\mathbf{k}\!\cdot\!\mathbf{p}$ theory for $\Gamma$-states}

The energy levels and wavefunctions of zone-center electron and hole
states are calculated using the eight-band $\mathbf{k}\!\cdot\!\mathbf{p}$
model, which was originally developed for the description of electronic
states in bulk materials~\cite{END95,Pol90,End95c}. 
In the context of heterostructures, the envelope function version
of the model has been applied to quantum wells~(QWs)~\cite{GHB93},
quantum wires~\cite{StB97}, and QDs~\cite{JIA97,Pry98,SGB99a,MBT04,WSB06}.
Details of the principles of our implementation are outlined in Ref.~\cite{StB97,WSB06}.

This model enables us to treat QDs of arbitrary shape and material
composition, including the effects of strain, piezoelectricity, VB
mixing, and CB-VB interaction. The strain enters our model via deformation
potentials as outlined by Bahder~\cite{Bah90}. Its impact on the local bandedges as a function of the QD geometry will be discussed further below. 

Due to the limited number of Bloch functions used for the wavefunction
expansion, the results of the eight-band $\mathbf{k}\!\cdot\!\mathbf{p}$
model are restricted to close vicinity of the Brillouin zone center.
However, as mentioned before, we calculate off-center states using the effective
mass model, detailed in the next paragraph.

\subsubsection{Effective mass theory for L- and X-states}

The single-particle states for L- and X-electrons are obtained within
the envelope function method based on effective mass approximation,~i.e.,
the following equation is solved~\cite{t_zibold} 
\begin{equation}
\hat{H}^{\mathrm{L,X}}F({\pmb r})=EF({\pmb r}),\label{Eq:singleBandEnvel-1}
\end{equation}
where $E$ and $F({\pmb r})$ are the eigen energy and the envelope
function, respectively, and $\hat{H}^{\mathrm{L,X}}$ is given by
\begin{equation}
\hat{H}^{\mathrm{L,X}}=-\frac{\hbar^{2}}{2}{\pmb\nabla}\cdot\left(\frac{1}{\underline{m}^{*}({\pmb r})}\right){\pmb\nabla}+E_{c}^{\mathrm{L,X}}({\pmb r})+V_{\mathrm{ext}}({\pmb r}).\label{Eq:singleBandEnvelHamiltonian-1}
\end{equation}
Here, $E_{c}^{\mathrm{L,X}}({\pmb r})$ is the positionally dependent
bulk conduction band energy for L or X point, $V_{\mathrm{ext}}({\pmb r})$
is the external potential induced by,~e.g., elastic strain, and $\pmb\nabla\equiv\left(\frac{\partial}{\partial x},\frac{\partial}{\partial y},\frac{\partial}{\partial z}\right)^{T}$
is the gradient. The effective mass parameter $\underline{m}^{*}({\pmb r})$
is given by~\cite{t_zibold} 
\begin{equation}
\underline{m}^{*}({\pmb r})=\left[m_{l}^{*}({\pmb r})-m_{t}^{*}({\pmb r})\right]\hat{\pmb k}_{0}\hat{\pmb k}_{0}^{T}+m_{t}^{*}({\pmb r}){\pmb1}_{3x3},\label{Eq:LXeffectiveMass-1}
\end{equation}
where $m_{l}^{*}({\pmb r})$ and $m_{t}^{*}({\pmb r})$ are positionally
dependent longitudinal and transversal effective masses, respectively,
$\hat{\pmb k}_{0}=\left<100\right>$ ($\hat{\pmb k}_{0}=\left<111\right>/\sqrt{3}$)
for X-point (L-point) of the Brillouin zone and ${\bf 1}_{3 \times 3}$ is
$3 \times 3$ the identity matrix.



\subsubsection{Strain and its effect on local bandedges}

As the impact of strain on the confinement is comparable to the band
offsets at the heterojunctions, the wavefunctions and energies are
very sensitive to the underlying strain distribution. The natural
choice of appropriate strain model in the context of multiband $\mathbf{k}\cdot\mathbf{p}$
theory is the continuum elasticity model \cite{Grundmann:95}. Its pros and
cons compared to valence-force-field like models are discussed in a
number publications \cite{SGB99a,Schliwa2007,PKW98b}. The magnitude of the strain
induced band-shifts is determined by the material dependent deformation
potentials \cite{BaS50b,herring_transport_1956}. For the CB
$\Gamma$-point, as well as for the valleys at the X-point and the
L-point, the strain induced energy shift is given by \cite{herring_transport_1956}:
\begin{equation}
E_{\mathrm{c}}^{i}(\hat{\boldsymbol{k}}_{0},\varepsilon)=E_{\mathrm{c}}^{i}(\hat{\boldsymbol{k}}_{0})+\Xi_{d}^{i}tr(\varepsilon)+\Xi_{u}^{i}(\hat{\boldsymbol{k}}_{0}\cdot\varepsilon\hat{\boldsymbol{k}}_{0})\quad\label{eq:Herring}
\end{equation}
with the absolute $\Xi_{d}^{i}$ and the uniaxial
$\Xi_{u}^{i}$ deformation potentials, $i\in\{\Gamma,\,{\rm L},\,{\rm X}\}$; $\varepsilon$ is the strain.

Evaluating Eq.~(\ref{eq:Herring}) for the strain conditions at the
vertical centerline of our QD with $\varepsilon_{xy}=\varepsilon_{xz}=\varepsilon_{yz}=0$
one arrives at:
\begin{eqnarray*}
	E_{\mathrm{c}}^{\Gamma}([000],\varepsilon) & = & E_{\mathrm{c}}^{\Gamma}+a_{c}^{\Gamma}tr(\varepsilon)\quad,\\
	E_{\mathrm{c}}^{L}([111],\varepsilon) & = & E_{\mathrm{c}}^{L}+a_{c}^{L}tr(\varepsilon)+\frac{1}{3}a_{cu}^{L}(\varepsilon_{xx}+\varepsilon_{yy}+\varepsilon_{zz})\quad,\\
	E_{\mathrm{c}}^{X}([100],\varepsilon) & = & E_{\mathrm{c}}^{X}+a_{c}^{X}tr(\varepsilon)+a_{cu}^{X}(\varepsilon_{xx})\quad,\\
	E_{\mathrm{c}}^{X}([010],\varepsilon) & = & E_{\mathrm{c}}^{X}+a_{c}^{X}tr(\varepsilon)+a_{cu}^{X}(\varepsilon_{yy})\quad,\\
	E_{\mathrm{c}}^{X}([001],\varepsilon) & = & E_{\mathrm{c}}^{X}+a_{c}^{X}tr(\varepsilon)+a_{cu}^{X}(\varepsilon_{zz})\quad.
\end{eqnarray*}
where $a_{\mathrm{c}}$ being the absolute deformation potential and $a_{\mathrm{cu}}$
the uniaxial shear deformation potential in $[100]$-direction of
CB.

The expression for $E_{\mathrm{c}}^{L}([111],\varepsilon)$ is identical
for all $L$-points, whereas a strain dependent splitting occurs between
the energies of $E_{\mathrm{c}}^{X}([100],\varepsilon)$, $E_{\mathrm{c}}^{X}([010],\varepsilon)$
and $E_{\mathrm{c}}^{X}([001],\varepsilon)$. At the QD's centerline,
however, $\varepsilon_{xx}=\varepsilon_{yy}$ holds and the course
of $E_{\mathrm{c}}^{X}([100],\varepsilon)$, $E_{\mathrm{c}}^{X}([010],\varepsilon)$
is identical (see Fig.~\ref{fig:QDbands}). 

For VB the coupling between light-hole and split-off
band results in more complex expressions \cite{chao_spin-orbit-coupling_1992}. With $\delta E=\frac{1}{2}a_{ub}(\varepsilon_{xx}+\varepsilon_{yy}-2\varepsilon_{zz})$
one obtains:
\begin{eqnarray}
E_{\mathrm{v}}^{\mathrm{HH}}(\Gamma,\varepsilon) & = & E_{\mathrm{v}}^{\Gamma}+a_{v}tr(\varepsilon)-\delta E\quad,\nonumber \\
E_{\mathrm{v}}^{\mathrm{LH}}(\Gamma,\varepsilon) & = & E_{\mathrm{v}}^{\Gamma}+a_{v}tr(\varepsilon)+\frac{1}{2}(\delta E-\Delta_{SO})\nonumber\\
& + & \frac{1}{2}(\sqrt{\Delta_{SO}^{2}+2\Delta_{SO}\cdot\delta E+9\delta E^{2}})\quad,\label{eq:LH}\\
E_{\mathrm{v}}^{\mathrm{SO}}(\Gamma,\varepsilon) & = & E_{\mathrm{v}}^{\Gamma}-\Delta_{\mathrm{SO}}+a_{v}tr(\varepsilon)+\frac{1}{2}(\delta E+\Delta_{SO})\nonumber\\
& - & \frac{1}{2}\sqrt{\Delta_{SO}^{2}+2\Delta_{SO}\cdot\delta E+9\delta E^{2}})\quad,\label{eq:SO}
\end{eqnarray}
with $a_{v}$ being the absolute deformation potential and $a_{ub}$
the uniaxial shear deformation potential in $[100]$-direction of
VB. $\Delta_{\mathrm{SO}}$ denotes the spin-orbit splitting
and $E_{\mathrm{v}}^{\Gamma}$ the energy of the unstrained valence
bandedge. 

Remarkably, there is a large coupling of light-hole and split-off
band (through the term $2\Delta_{SO}\cdot\delta E$ under the root
of Eq.~(\ref{eq:LH}) owing to both a sizable spin-orbit coupling,
$\Delta_{\mathrm{SO}}$, and a large biaxial strain leading to large
values of $\delta E$. As a result, the light-hole band becomes upshifted
by at least 100 meV within the QD. 

We would like to stress that the aim of the above analysis of strain-induced energy shifts was to show the general trends affecting,~e.~g., the computation of bandedges. Calculating single-particle states of our QDs we evaluated $E_{\mathrm{c}}(\hat{\boldsymbol{k}}_{0},\varepsilon)$, $E_{\mathrm{v}}^{\mathrm{HH}}(\Gamma,\varepsilon)$, $E_{\mathrm{v}}^{\mathrm{LH}}(\Gamma,\varepsilon)$, and $E_{\mathrm{v}}^{\mathrm{SO}}(\Gamma,\varepsilon)$ in each point of the simulation space and included the effects of shear strain.

\subsubsection{Piezoelectricity}

Piezoelectricity is defined as the generation of electric polarization
by the application of stress to a crystal lacking a center of symmetry~\cite{CAD46}.  
Following our previous works \cite{SGB99a,Schliwa:09,Klenovsky2018}, we calculate
the piezoelectric polarization (${\bf P}$) in first (${\bf P}_{l}$) and second (${\bf P}_{nl}$) order~\cite{Bester:06,Beya-Wakata2011}


%
\begin{equation}
\label{eq:1stPiez}
{\bf P}_{l}=e_{14}\begin{pmatrix}\varepsilon_4\\\varepsilon_5\\\varepsilon_6\end{pmatrix},
\end{equation}
and 
%
%
\begin{equation}
\label{eq:2ndPiez}
{\bf P}_{nl}=B_{114}\begin{pmatrix}\varepsilon_1\varepsilon_4\\\varepsilon_2\varepsilon_5\\\varepsilon_3\varepsilon_6\end{pmatrix}+
B_{124}\begin{pmatrix}\varepsilon_4(\varepsilon_2+\varepsilon_3)\\\varepsilon_5(\varepsilon_3+\varepsilon_1)\\\varepsilon_6(\varepsilon_1+\varepsilon_2)\end{pmatrix}+
B_{156}\begin{pmatrix}\varepsilon_5\varepsilon_6\\\varepsilon_4\varepsilon_6\\\varepsilon_4\varepsilon_5\end{pmatrix},
\end{equation}
where $\varepsilon_i$ are indexed according to the Voigt notation,~i.e.,~ $\varepsilon_1\equiv\varepsilon_{xx}$, $\varepsilon_2\equiv\varepsilon_{yy}$, $\varepsilon_3\equiv\varepsilon_{zz}$, $\varepsilon_4\equiv2\varepsilon_{yz}$, $\varepsilon_5\equiv2\varepsilon_{xz}$, $\varepsilon_6\equiv2\varepsilon_{xy}$,~\cite{Beya-Wakata2011} where $x,y,z$ denote the crystallographic axes of the conventional cubic unit cell of the zincblende lattice. The values of the parameters $e_{14}$, $B_{114}$, $B_{124}$, and $B_{156}$ are given in Ref.~\cite{SupMatPRBgen}.

The resulting piezoelectric potential is obtained by solving the Poisson\textquoteright s equation, taking into account the material dependence of the static dielectric constant $\ensuremath{\epsilon_{\mathrm{s}}(\mathbf{r})}$.

\subsection{Coulomb interaction}
As soon as more than one charge carrier is confined inside the QD, the influence of direct Coulomb interaction, exchange effects, and correlation lead to the formation of distinct multiparticle states which are calculated using the CI method. This method rests on a basis expansion of the excitonic Hamiltonians into Slater determinants, which consist of antisymmetrized products of single-particle wavefunctions, obtained from \ekp\, theory for $\Gamma$-point states.
The method is applicable within the strong confinement regime as the obtained basis functions are already similar to the weakly correlated many-body states \cite{BLS00,SFZ01,SSH01,Klenovsky2017}.

We proceed by giving a brief overview of the CI method used in this work, following Ref.~\cite{Klenovsky2017}. In CI we solve the stationary Schr\"{o}dinger equation
\begin{equation}
\label{Eq:CISchroedinger}
\hat{H}^M\left|M\right>=E^M\left|M\right>,
\end{equation}
where $E^M$ is the eigen energy of the (multi-)excitonic state $\left|M\right>$ corresponding to $N_a$ and $N_b$, i.e.,~the numbers of particles $a$ and $b$, respectively, where $a,b\in\{e,h\}$ with $e$ and $h$ standing for electron and hole, respectively. We look for solutions of Eq.~(\ref{Eq:CISchroedinger}) in the form
\begin{equation}
\label{eq:CIWavefunction}
\left|M\right>=\sum_{\nu}\eta_{\nu}\left|D^M_{\nu}\right>
\end{equation}
where $\nu$ runs over all $e$ and $h$ configurations in given $M$. The configurations are assembled in the form of the Slater determinants $\left|D^M_{\nu}\right>$, which are constructed from the single-particle basis states. Using the ansatz~(\ref{eq:CIWavefunction}) we obtain the coefficients $\eta_{\nu}$ by the variational procedure,~i.e.,~we solve the system of equations $\sum_m\left<D^M_n\right|\hat{H}^M\left|D^M_m\right>\eta_m=E^M\eta_n$, under the constraint $\sum_{\nu} |\eta_{\nu}|^2=1$.

The elements of the CI Hamiltonian are $\hat{H}^M_{nm} \equiv \langle D_n^M|{\hat{H}^M}|D_m^M\rangle  = \langle D_n^M|{\hat{H}_0^M}|D_m^M\rangle+\langle D_n^M|{\hat{V}^M}|D_m^M\rangle$. Here $\langle D_n^M|{\hat{H}_0^M}|D_m^M\rangle$ corresponds to the non-interacting (single-particle) part and the latter term introduces the Coulomb interaction of the kind
\begin{equation}
\label{Eq:CIcoulombFull}
\begin{split}
&\langle D_n^M|{\hat{V}^M}|D_m^M\rangle=\frac{1}{4\pi \epsilon_0}\sum_{ijkl}\iint {\rm d}{\bf r}_1 {\rm d}{\bf r}_2\frac{q_iq_j}{\epsilon({\bf r}_1, {\bf r}_2) |{\bf r}_1 - {\bf r}_2|} \times \\
&\{\psi^*_{i}({\bf r}_{1}) 
\psi^*_{j}({\bf r}_{2})\psi_{k}({\bf r}_{1})  \psi_{l}({\bf r}_{2})-\psi^*_{i}({\bf r}_{1}) 
\psi^*_{j}({\bf r}_{2})\psi_{l}({\bf r}_{1})  \psi_{k}({\bf r}_{2})\}\\
&= \sum_{ijkl}\left(V_{ij,kl} - V_{ij,lk}\right),\\
\end{split}
\end{equation}
where $\epsilon_0$ and $\epsilon({\bf r}_1, {\bf r}_2)$ are the vacuum and spatially varying relative dielectric constants, respectively, $q_i,q_j\in\{-e,+e\}$ where $e$ is the elementary charge, and the spatial position of the charges is marked by ${\bf r}_1$ and ${\bf r}_2$, respectively. The Coulomb interaction described by $V_{ij,kl}$ ($V_{ij,lk}$) is called direct (exchange).

We add a comment about an ongoing discussion~\cite{Bester2008,Benedict2002} related to the nature of the dielectric screening in Eq.~(\ref{Eq:CIcoulombFull}),~i.e., whether or not to set $\epsilon({\bf r}_1, {\bf r}_2)=1$ for the exchange integral, for both $V_{ij,kl}$ and $V_{ij,lk}$, or use bulk values in both cases. We tested those options by computing the fine-structure splitting (FSS) of exciton and separately also the trion binding energies (TBE) relative to exciton, both using CI for typical InAs/GaAs QD (lens-shape, base diameter 20~nm, height 3~nm). We found that setting $\epsilon({\bf r}_1, {\bf r}_2)=1$ for $V_{ij,lk}$ resulted in a rather realistic values of both FSS and TBE for the basis composed solely of the ground state electron and hole states. However, for larger basis, while the values of FSS remained within experimentally realistic limits~\cite{Luo2009}, those for TBE were found unreasonably large and were increasing with basis size without reaching saturation, when higher energy single-particle states were included in the basis. On the other hand, the CI results, when $\epsilon({\bf r}_1, {\bf r}_2)$ was set to bulk values for both $V_{ij,kl}$ and $V_{ij,lk}$, led to values of FSS and TBE within experimentally realistic limits, regardless of the CI basis size. Thus, on the grounds of inconsistent results obtained for $\epsilon({\bf r}_1, {\bf r}_2)=1$ we decided to use the bulk values of $\epsilon({\bf r}_1, {\bf r}_2)$ for both $V_{ij,kl}$ and $V_{ij,lk}$.

We finally note that the numerical difficulty connected with the evaluation of the six-fold integral in Eq.~(\ref{Eq:CIcoulombFull}) has been overcome using the Green's function method~\cite{Schliwa:09,t_stier},~i.e.:
\begin{equation}
\label{Eq:CIGreen}
\begin{split}
&\nabla\left[\epsilon({\bf r})\nabla \hat{U}_{ajl}({\bf r})\right]=\frac{4\pi e^2}{\epsilon_0}\psi_{aj}^{*}({\bf r})\psi_{al}({\bf r}),\\
&V_{ij,kl}=
\langle\psi_{bi}|\hat{U}_{ajl}|\psi_{bk}\rangle,
\end{split}
\end{equation}
where 
$a,b\in{\{e,h\}}$ and $\nabla\equiv\left(\frac{\partial}{\partial x},\frac{\partial}{\partial y},\frac{\partial}{\partial z}\right)^{\rm T}$.

\subsection{Optical properties}
The interband absorption and emission spectra are calculated by the Fermi's golden rule, see also Eq.~(\ref{Eq:KaneParamDirect}), applied to excitonic states calculated by the CI method, see Ref.~\cite{Klenovsky2017} for details. In this paper we focus on $\Gamma$-point transitions only, and leave the other results for a separate publication. This is motivated (i) by the discussion following Eq.~(\ref{Eq:KaneParamIndirect}) and by experiments presented in Ref.~\cite{ArxivSteindl:19}, where we report dominant contribution of $\Gamma$-point transitions in photoluminescence (PL) spectra of In$_{1-x}$Ga$_{x}$As$_{y}$Sb$_{1-y}$/GaAs/GaP QDs.

The radiative rates $R\equiv\Gamma_{fi}$ and transition of the considered $\Gamma$-point excitonic transitions are calculated according to
\begin{eqnarray}
R\equiv\Gamma_{fi} & = & \left(\frac{e}{m}\right)^{2}\frac{2\hbar\omega}{c^{3}}\left|\langle f|\boldsymbol{e}\cdot\hat{P}|i\rangle\right|^{2}\quad {\rm with}\label{Eq:CItransition}\\
\hat{P} &=& \sum_{n,m}\langle\psi_f^n|\nabla|\psi_i^m\rangle\quad,\label{Eq:spCItransitionSum}\\
\langle\psi_f^n|\nabla|\psi_i^m\rangle & = & \sum_{j,k}^{8}\langle F_{j}u_{j}^{\Gamma}|\nabla|F_{k}u_{k}^{\Gamma}\rangle\label{Eq:8kptransition}\\
& = & \sum_{j,k}^{8}[\delta_{jk}\langle F_{j}|\nabla|F_{k}\rangle+\langle F_{j}|F_{k}\rangle\langle u_{j}^{\Gamma}|\nabla|u_{k}^{\Gamma}\rangle] \quad,\nonumber
\end{eqnarray}
%
%
%
where $e$ and $m$ are the elementary charge and the mass of the free electron, respectively, $\hbar\omega$ is the energy of the emitted radiation with $\hbar$ being the reduced Planck's constant and $\omega$ the angular frequency of the radiation, respectively. Furthermore, $|i\rangle$ and $\langle f|$ ($|\psi_i\rangle$ and $\langle\psi_f|$) mark the initial and final multi- (single-)particle state, respectively, $F_{j}$ denotes the envelope function, $u_{j}^{\Gamma}$ the associated Bloch function with band indexes $j$, $k$, and $\boldsymbol{e}$ is the polarization vector; $\delta_{jk}$ is the Kronecker symbol. We dropped the indices $m$ and $n$ on the right hand side of Eq.~(\ref{Eq:8kptransition}) because of no risk of confusion. Note, that in Eq.~(\ref{Eq:CItransition}) the inner product $\boldsymbol{e}\cdot\hat{P}$ must be performed before projecting $\hat{P}$ on CI states and the summation in Eq.~(\ref{Eq:spCItransitionSum}) runs over single-particle states $\psi_f^n$ and $\psi_i^m$ present in CI states $\langle f|$ and $|i\rangle$, respectively.
%
%
%
%
%

\section{Confinement potentials}

\begin{figure}[!ht]
	\begin{center}
		\begin{tabular}{c}
			\includegraphics[width=0.48\textwidth]{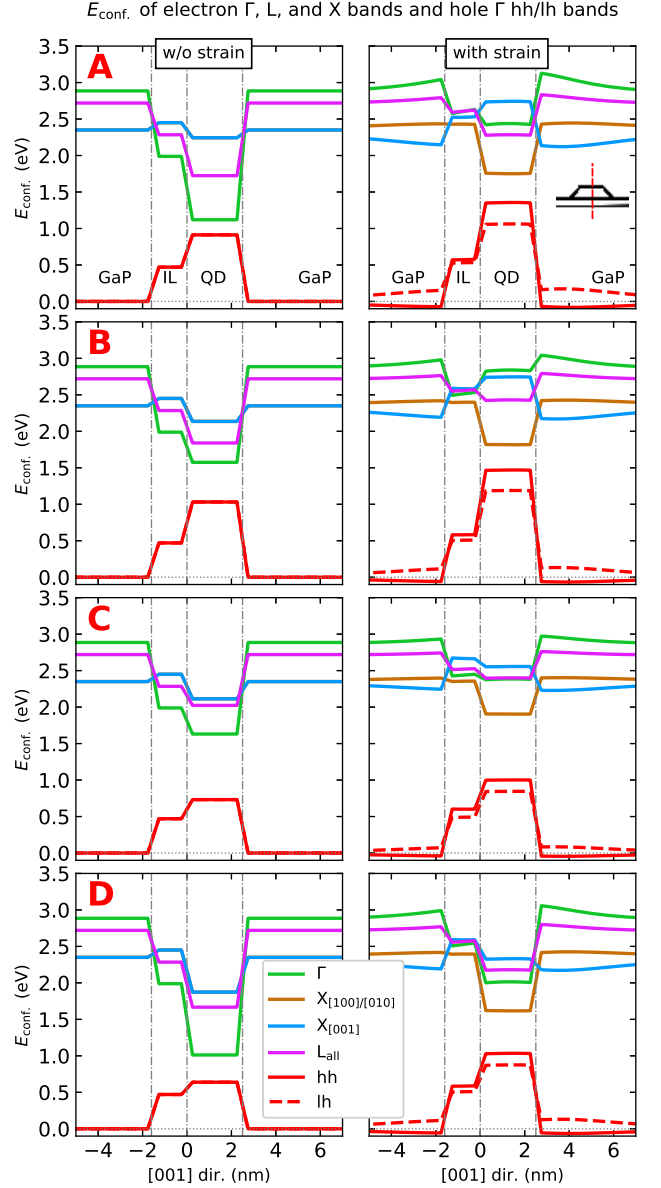} 
		\end{tabular}
	\end{center}
	\caption{$E_{\rm conf.}$ of electrons for several points of ${\bf k}$-space ($\Gamma$, X, and L) and of holes at $\Gamma$, given along [001] crystal direction along QD vertical symmetry axis.  We show on the left (right) column $E_{\rm conf.}$ without (with) considering the strain field in and around the QD. The insets show the designation of bands and the sketch of the direction where the evaluation of $E_{\rm conf.}$ was performed with respect to Fig.~\ref{fig:QDsketch}~(c). Note, that we show only X$_{[100]/[010]}=1/2\times($X$_{[100]}+$X$_{[010]})$ and L$_{\mathrm{all}}=1/4\times\sum_{i=1}^4$L$_{i}$ here, see text for details. The capital letter A, B, C, and D mark the concentrations listed in Tab.~\ref{tab:ABCDdesig}.
		\label{fig:QDbands}}
\end{figure}
We start with the single-particle confinement potentials ($E_{\rm conf.}$) for electrons and holes and show the results in Fig.~\ref{fig:QDbands} for $E_{\rm conf.}$ along the QD growth axis parallel to [001] crystal direction, computed without and with the inclusion of elastic strain. We firstly notice that the strain has considerable effect on $E_{\rm conf.}$ except for X$_{[100]/[010]}$ states which are bound inside QD body and for which $E_{\rm conf.}$ attains the lowest energy in our structure, similarly to (In,Ga)As/GaP QDs~\cite{Robert2012,Robert2014,Robert2016}. On the other hand, the bands which are influenced much more by strain are X$_{[001]}$ and particularly $\Gamma$, for which the strain can even revert the position of the minimum of $E_{\rm conf.}$ outside of QD body. For the former (X$_{[001]}$), the minimum of $E_{\rm conf.}$ occurs above QD due to the tensile $\epsilon_{zz}$ strain exerted by the dot body. We note that similar effect occurs also in SiGe/Si~\cite{KlenovskyPRB2012} and (In,Ga)As/GaP~\cite{Robert2014} QD systems. For the latter ($\Gamma$) the minimum is found in the GaAs-IL for Sb rich dots. 
As shown in Ref.~\cite{t_sala}, during the growth an Sb-soaking after the GaAs-IL deposition is employed prior to QD-nucleation. This is very likely to trigger an As-for-Sb anion exchange reaction at the GaAs-IL surface, leading to GaSb formation and thus a considerable material intermixing in the QD layer. Therefore, such intermixing leads there to the minimum of $E_{\rm conf.}$ for 
$\Gamma$-electrons ($E^{c,\Gamma}_{\rm conf.}$) to be strongly positionally dependent in In$_{1-x}$Ga$_x$As$_y$Sb$_{1-y}$/GaAs/GaP QDs. Finally, we note that $E_{\rm conf.}$ for L bands are affected by a mere increase in energy.

\section{Results for single-particle states}

\begin{figure}[!ht]
	\begin{center}
		\begin{tabular}{c}
			\includegraphics[width=0.48\textwidth]{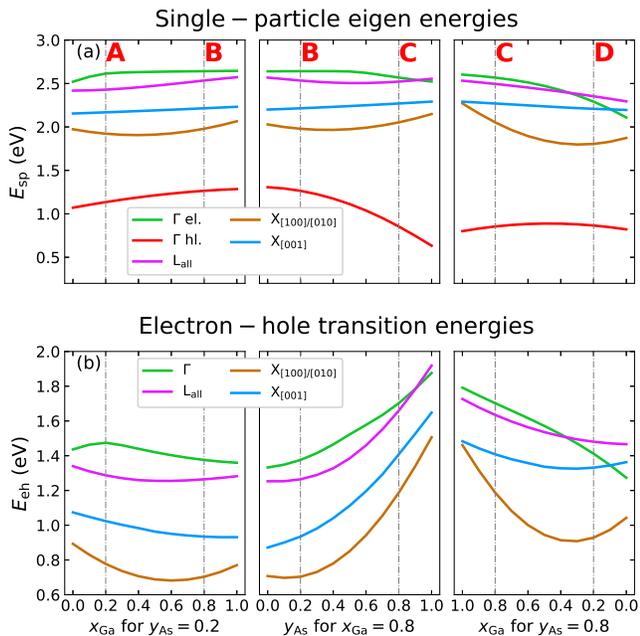}
		\end{tabular}
	\end{center}
	\caption{(a) Ground state single-particle eigen~energy ($E_{\rm sp}$) and  (b) electron-hole transition energy ($E_{\rm eh}$) for selected alloys as described in Tab.~\ref{tab:ABCDdesig}. For L-~and~X$_{[100]/[010]}$ electrons the plotted energies are averaged over the first eight almost degenerate $L$-levels labelled L$_{\mathrm{all}}$ and over the first four almost degenerate X-levels, denoted with X$_{[100]/[010]}$, respectively.
		\label{fig:QDexcitonEF}}
\end{figure}

We now proceed with the results for  single-particle states of our In$_{1-x}$Ga$_x$As$_y$Sb$_{1-y}$/GaAs/GaP QDs. For the alloys listed in table~\ref{tab:ABCDdesig} we show the results for~$\Gamma$-, L-,~and~X-electron and $\Gamma$-hole ground state energies ($E_{\rm sp}$) and the related interband transition energies ($E_{\rm eh}$) in Fig.~\ref{fig:QDexcitonEF}. First of all, we observe that the first eight states involving L-electrons are almost degenerate in energy, hence, we do not distinguish between them in Fig.~\ref{fig:QDexcitonEF} and group them under the label L$_{\rm all}$. The same holds true for (X$_{[100]}$, X$_{010]}$) electrons, which we denote X$_{[100]/[010]}$. Interestingly, $E_{\rm sp}$ for $\Gamma$-electron states crosses that for L$_{\rm all}$ close to point C in the middle panel of Fig.~\ref{fig:QDexcitonEF}~(a) and both L$_{\rm all}$ and X$_{[001]}$ close to point D in the rightmost panel of Fig.~\ref{fig:QDexcitonEF}~(a).

However, since $E_{\rm sp}$ of electrons does not change considerably with dot composition, $E_{\rm eh}$ between electrons and holes is dictated by $E_{\rm sp}$ of the latter, see Fig.~\ref{fig:QDexcitonEF}~(b). 
The energy $E_{\rm sp}$ of holes is mainly influenced by antimony content which is, indeed, one of the main features of our QD system and it will be important also when using of our dots for information storage in QD-Flash memory and for the quantum gate proposal are discussed later. The energies of holes, thus, cause the large increase in $E_{\rm eh}$ of~$\sim$500~meV for recombinations between $\Gamma$-electron to $\Gamma$-hole states or even up to~$\sim$700~meV for transitions from X$_{[100]/[010]}$, see middle panel of Fig.~\ref{fig:QDexcitonEF}~(b).
On the other hand, $E_{\rm sp}$ of electrons dictates the energy ordering of $E_{\rm eh}$ which is for most Ga and As concentrations from highest to lowest: $\Gamma$, L$_{\mathrm{all}}$, X$_{[001]}$, and X$_{[100]/[010]}$, see Fig.~\ref{fig:QDexcitonEF}~(b). This is also the case for the energy flipping of $E_{\rm eh}$ for transitions from $\Gamma$ and L$_{\rm all}$ or X$_{[001]}$ to $\Gamma$ holes.

\begin{figure}[!ht]
	\begin{center}
		\begin{tabular}{c}
			\includegraphics[width=0.48\textwidth]{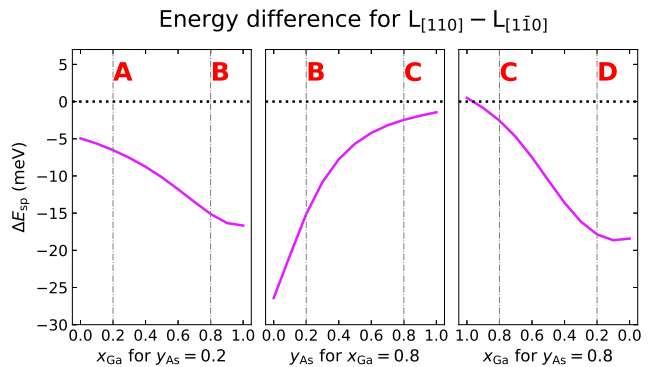}
		\end{tabular}
	\end{center}
	\caption{Energy difference $\Delta E_{\rm sp}$ between ${\rm L_{[110]}}$ and ${\rm L_{[1-10]}}$ electrons, $\Delta E_{\rm sp}=E^{[110]}_{\rm sp}-E^{[1-10]}_{\rm sp}$ for selected alloys as described in Tab.~\ref{tab:ABCDdesig}.
		\label{fig:QDLfine}}
\end{figure}
For completeness, we find the energy difference ($\Delta E_{\rm sp}$) between ${\rm L_{[110]}}$ and ${\rm L_{[\bar{1}\bar{1}0]}}$, ${\rm L_{[\bar{1}10]}}$ and ${\rm L_{[1\bar{1}0]}}$, and ${\rm X_{[100]}}$ and ${\rm X_{[010]}}$ electrons to be smaller than $1\,{\rm \mu eV}$ in our structure. However, $\Delta E_{\rm sp}$ attains values of several tens of meV when computed between ${\rm L_{[110]}}$ and ${\rm L_{[1\bar{1}0]}}$ bands, see Fig.~\ref{fig:QDLfine}. Clearly, $E_{\rm sp}$ for ${\rm L_{[110]}}$ electrons is smaller than for ${\rm L_{[1\bar{1}0]}}$, which is a result of the combined effect of shear strain, see Eq.~(\ref{eq:Herring}), and piezoelectricity, Eqs.~(\ref{eq:1stPiez})~and~(\ref{eq:2ndPiez}), for strained QDs fabricated from zincblende crystals due to their non-centrosymmetricity. The energy splitting seen in Fig.~\ref{fig:QDLfine} is computed without taking into account mixing of L-Bloch waves with other electron bands which should be, however, rather small~\cite{Robert2016}.

We note, that we were able to observe transitions like those shown in Fig.~\ref{fig:QDexcitonEF} by PL for two samples with In$_{0.2}$Ga$_{0.8}$As$_{0.8}$Sb$_{0.2}$/GaAs/GaP and In$_{0.5}$Ga$_{0.5}$As/GaAs/GaP QDs, respectively, in Ref.~\cite{ArxivSteindl:19}. A suitable method to observe transitions between ${\bf k}\neq 0$-electrons and $\Gamma$-holes, is a resonant PL technique similar,~e.~g.,~to that used in Ref.~\cite{Rautert2019} for study of (In,Al)As/AlAs QDs.

\newpage

\onecolumngrid

\begin{center}
	\begin{figure}[!ht]
		\begin{center}
			\begin{tabular}{c}
				\includegraphics[width=0.95\textwidth]{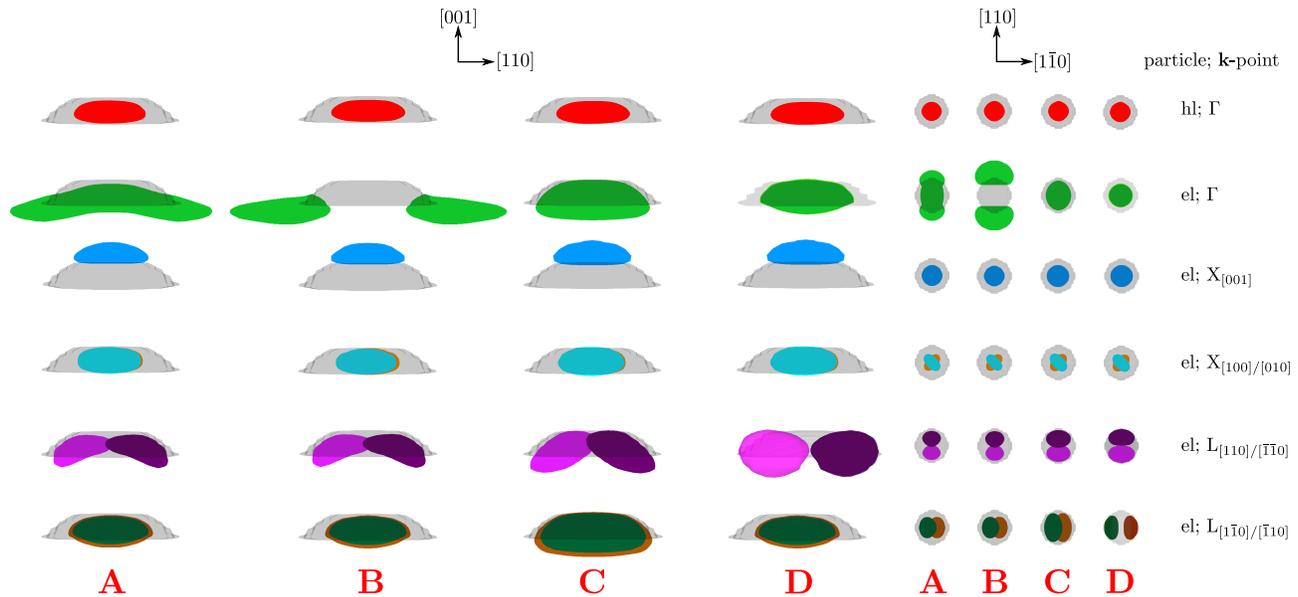} \\
			\end{tabular}
		\end{center}
		\caption{Side and top views of the probability densities of electrons [el] and holes [hl] in In$_{1-x}$Ga$_x$As$_y$Sb$_{1-y}$/GaAs/GaP QDs \{grey objects\}. Cuts through the plane parallel (perpendicular) to the QD symmetry axis are given from the first to the fourth (fifth to eight) column for $x_{\rm Ga}$ and $y_{\rm As}$ corresponding to A, B, C, and D in Tab.~\ref{tab:ABCDdesig}. The designation of the quasi-particles and the corresponding Bloch waves are given in the ninth column. The properties of QD is the same as that in Fig.~\ref{fig:QDsketch}~(c).
			The single-particle electron and hole envelope functions for $\Gamma$-point Bloch states are calculated using eight-band ${\bf k}\cdot{\bf p}$, those for X- and L-electron states by effective mass theory (see main text).
			The isosurfaces encircle 90~\% of total probability density.
			Due to the $\bf{k}$-space (a)symmetry, some of the X and L states for QD with circular base are almost degenerate in (001) plane, thus, we group them together in the lower three rows of the figure. In the top row of the figure we show the crystallographic orientations to facilitate the comparison with the orientation of the probability densities.
			\label{fig:QDwfs}}
	\end{figure}
\end{center}

\twocolumngrid

\begin{center}
	\begin{figure}[!ht]
		\begin{center}
			\begin{tabular}{c}
				\includegraphics[width=0.48\textwidth]{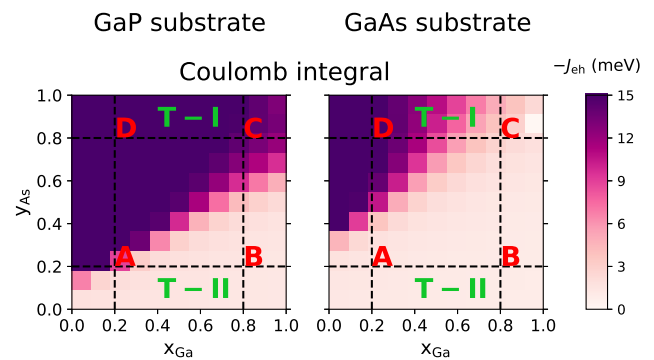} \\
			\end{tabular}
		\end{center}
		\caption{Electron-hole Coulomb integral ($-J_{\rm eh}$) for In$_{1-x}$Ga$_x$As$_y$Sb$_{1-y}$ QD with GaAs IL grown on GaP (left) and on GaAs (right) substrates, respectively. Except for the composition, QD structural properties were the same as those in Fig.~\ref{fig:QDsketch}~(c). The capital letters A, B, C, and D mark the concentrations listed in Tab.~\ref{tab:ABCDdesig}. For the alloy interpolation scheme see Eqs.~(\ref{Eq:alloy_lin-1})~and~(\ref{Eq:alloy_tern-1}). The marks T-I and T-II denote the type of confinement in real space.
			\label{fig:QDgammaDirCoul}}
	\end{figure}
\end{center}

We proceed with the inspection of the wavefunctions of In$_{1-x}$Ga$_x$As$_y$Sb$_{1-y}$/GaAs/GaP QDs for $x_{\rm Ga}$ and $y_{\rm As}$ corresponding to A, B, C, and D (see Tab.~\ref{tab:ABCDdesig}), and we show that in Fig.~\ref{fig:QDwfs}. We find that the spatial location of the probability densities of states confirms our expectations drawn from the inspection of $E_{\rm conf.}$ in Fig.~\ref{fig:QDbands}. In particular, it allows us to make an assignment of the type of confinement of the $\Gamma$-electrons in real space. Thus, C and D contents seem to correspond to type-I transition of $\Gamma$-electrons to $\Gamma$-holes, while B is type-II, and A corresponds to the transition between those two types of confinement. Further, the spatial position of wavefunctions shows that transitions involving X$_{[001]}$-electrons are of type-II, and those for L$_{\rm all}$ and X$_{[100]/[010]}$ of type-I nature in real space, regardless of $x_{\rm Ga}$ and $y_{\rm As}$ contents in the dot.

However, the assignment of $\Gamma$-transitions can be done more precisely based on the inspection of the corresponding electron-hole Coulomb integrals ($-J_{ \rm eh}$), see Fig.~\ref{fig:QDgammaDirCoul}.
We see that $-J_{\rm eh}$ is by far smaller for type~II compared to type~I, owing to the spatial separation of the quasiparticles. Clearly, type~I occurs in our system for dots rich in Indium and Arsenic, while those with larger Ga and Sb tend to be type-II.
Notice also the comparison between GaP and GaAs substrates in Fig.~\ref{fig:QDgammaDirCoul}. We will return to the identification of the type of confinement from the properties of excitons in the following.

\subsection{Hole localization energies and storage time}

\begin{figure}[!ht]
	\begin{center}
		\begin{tabular}{c}
			\includegraphics[width=0.48\textwidth]{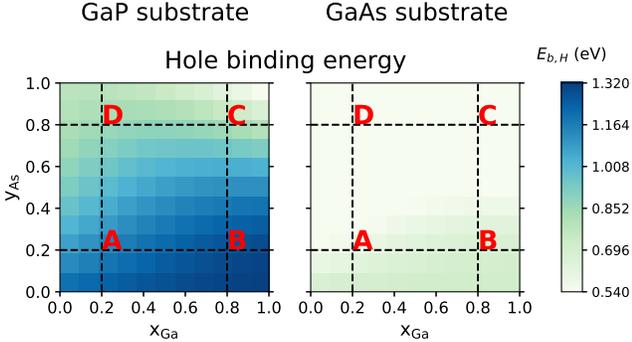}
		\end{tabular}
	\end{center}
	\caption{Binding energies of single-particle hole states ($E_{b,H}$) for In$_{1-x}$Ga$_x$As$_y$Sb$_{1-y}$ QD with GaAs IL grown on GaP (left column) or on GaAs (right column) substrates. The single-particle hole energies ($E_{H}$) necessary for the computation of $E_{b,H}$ (see text) were obtained within the envelope approximation based on eight-band ${\bf k}\cdot{\bf p}$ method. Notice that dots grown on GaP provide more than twice larger $E_{b,H}$ than those on GaAs. Except for the composition, the QD structural properties were the same as those in Fig.~\ref{fig:QDsketch}~(c). The capital letter A, B, C, and D mark the concentrations listed in Tab.~\ref{tab:ABCDdesig}. For the alloy interpolation scheme see Eqs.~(\ref{Eq:alloy_lin-1})~and~(\ref{Eq:alloy_tern-1}).
		\label{fig:HoleConf}}
\end{figure}

The variations of the QD valence bandedge energies upon chemical composition translates into a large variation of the hole localization energy defined by~\cite{t_marent,t_sala,Sala2018} $E_{b,H}=E_{H}-E^{v,\Gamma}_{\infty}$ with $E_{H}$ being energy of the single-particle hole state and $E^{v,\Gamma}_{\infty}$ the substrate material  $\Gamma$-VB energy, respectively. The results for $E_{b,H}$ are shown in Fig.~\ref{fig:HoleConf} as function of composition in In$_{1-x}$Ga$_x$As$_y$Sb$_{1-y}$ QDs on GaAs-IL grown either on GaP or GaAs substrates. Evidently, the QDs grown on GaP exhibit more than twice $E_{b,H}$ compared to QDs on GaAs, thus, confirming the importance of substrate material for QD-Flash concept~\cite{t_nowozin,Sala2016,t_sala,Sala2018}.

The energy $E_{b,H}$ can be translated into the storage time of QD-Flash memory units by using the expression~\cite{t_marent,t_nowozin,Marent2011,t_sala}
\begin{equation}
\label{eq:QDFlashStorageTime}
\tau=\frac{1}{\gamma\sigma_{\infty}T^2}\exp{\left(\frac{E_{b,H}}{k_{\rm B}T}\right)},
\end{equation}
with $\gamma=\sqrt{3(2\pi)^3}E_{H}m_v^*k_{\rm B}^2/h^3$ depending on the bulk material valence $\Gamma$-band effective mass $m_v^*$, $k_{\rm B}$ being the Boltzmann constant, $\sigma_{\infty}$  the capture cross-section, and $T$ the temperature. If we let $m_v^*$ to depend on $x_{\rm Ga}$ and $y_{\rm As}$ and choose $\sigma_{\infty}=9\times 10^{11}$~cm$^2$ from Ref.~\cite{t_sala,Sala2018} we find that the maximum $E_{b,H}=1.32$~eV in Fig.~\ref{fig:HoleConf} relates to a storage time of~5000~s, occurring for pure GaSb QD with GaAs-IL grown in GaP. However, $\sigma_{\infty}$ is a sensible parameter entering the calculation of the storage-time: it depends on the chemical composition and the QD-morphology itself (cmp. Ref.~\cite{bonato_230_2015}:$\sigma_{\infty}=(8 \pm 5)\times 10^{10}$~cm$^2$   and Ref.~\cite{Sala2018}:$\sigma_{\infty}=(9 \pm 5) \times 10^{11}$~cm$^2$). Both properties, $E_{b,H}$ and $\sigma_{\infty}$, are subject of constant technological optimization. Note that the value of $\sigma_{\infty}$ is not part of our modelling scheme but enters the calculation as external parameter~\cite{t_nowozin}.

\section{${\bf \Gamma}$-excitons}
We utilize the obtained single-particle wavefunctions as basis states for CI calculations and compute the corresponding exciton ($X^0$) states. Since we previously set $P_{\mathrm{{\bf k}\neq0}}=0$, it is reasonable to evaluate in the following $X^0$ consisting of $\Gamma$-electrons and $\Gamma$-holes only to avoid omission of some Coulomb elements for complexes involving ${\bf k}\neq 0$ electrons.

\begin{figure}[!ht]
\renewcommand{\tabcolsep}{2pt}
\begin{center}
\begin{tabular}{c}
\includegraphics[width=0.48\textwidth]{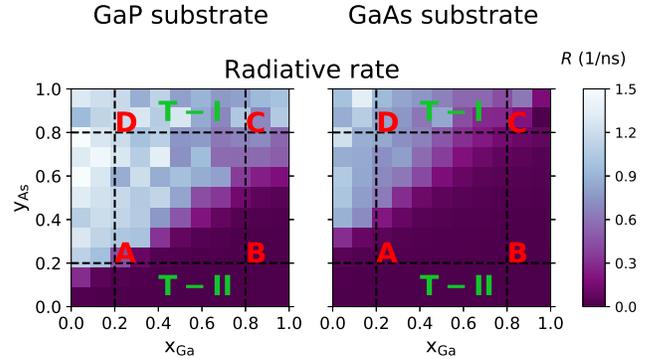}
\end{tabular}
\end{center}
\caption{Emission radiative rate ($R$) of bright $X^0$ for In$_{1-x}$Ga$_x$As$_y$Sb$_{1-y}$ QD with GaAs IL grown on GaP (left) and on GaAs (right) substrates, respectively. The single-particle basis of CI calculations was two electron and two hole ground states. Notice that type-I dots occur for larger $x_{\rm Ga}$ and $y_{\rm As}$ for QDs grown on GaP than on GaAs. Except for the composition QD structural properties were the same as those in Fig.~\ref{fig:QDsketch}~(c). The letters A, B, C, and D mark the concentrations listed in Tab.~\ref{tab:ABCDdesig}. For the alloy interpolation scheme see Eqs.~(\ref{Eq:alloy_lin-1})~and~(\ref{Eq:alloy_tern-1}). The marks T-I and T-II denote the type of confinement in real space.
\label{fig:QDexcitonRadRate}}
\end{figure}
We first discuss the emission radiative rate ($R$) of $X^0$ calculated using the Fermi's Golden rule as was discussed earlier, see also Ref.~\cite{Klenovsky2017} for details. The results for a number of $x_{\rm Ga}$ and $y_{\rm As}$ values are shown in Fig.~\ref{fig:QDexcitonRadRate}, and together with Fig.~\ref{fig:QDgammaDirCoul}, allow us to find the contents for which In$_{1-x}$Ga$_x$As$_y$Sb$_{1-y}$/GaAs/GaP QDs show type-I or type-II confinement.
Type I can be expected for $y_{\rm As}/x_{\rm Ga}\gtrsim 1$ and consequently type II for $y_{\rm As}/x_{\rm Ga}\lesssim 1$. We also show in Fig.~\ref{fig:QDexcitonRadRate} the values of $R$ for the same dots on GaAs substrate for comparison. As expected, type II is associated with the amount of GaSb in the QD structure as found also elsewhere~\cite{Klenovsky2010a,Klenovsky2010}.
Interestingly, type I for GaAs substrate occurs mostly for QDs with larger values of $y_{\rm As}$ than for GaP substrate. This is again a result of much increased hydrostatic strain in the latter case, since the GaP substrate provides a considerably larger confinement for quasiparticles than the former. The aforementioned hints to the conclusion that QD structures grown on GaP might perform even better in optoelectronic applications than those grown on GaAs substrates which are currently under study~\cite{Fischbach2017}.

\begin{figure}[!ht]
\renewcommand{\tabcolsep}{2pt}
\begin{center}
\begin{tabular}{c}
\includegraphics[width=0.48\textwidth]{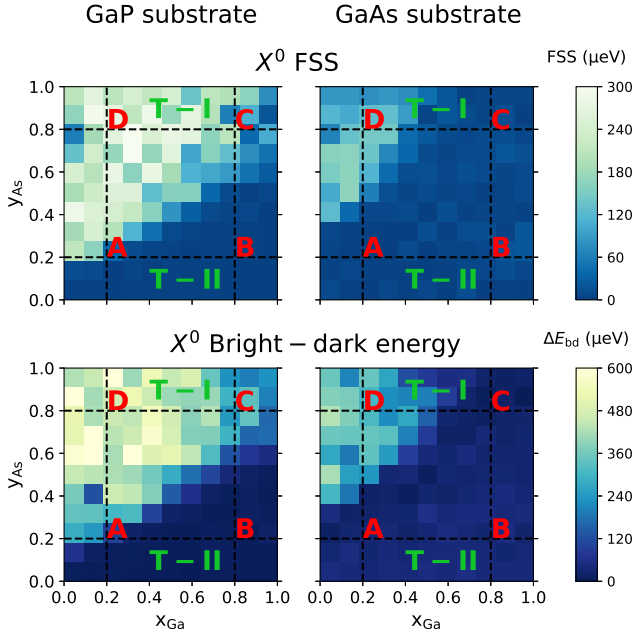}
\end{tabular}
\end{center}
\caption{Bright $X^0$ FSS and energy difference between bright and dark $X^0$ ($\Delta E_{\rm bd}$) for In$_{1-x}$Ga$_x$As$_y$Sb$_{1-y}$ QD with GaAs IL grown on GaP (left column) or on GaAs (right column) substrates, respectively. The single-particle basis of CI calculations was two electron and two hole ground states. Notice that FSS is generally larger for type-I dots grown on GaP than on GaAs. For both substrates type II is associated with very small FSS and $\Delta E_{\rm bd}$. Except for the composition, QD structural properties were the same as those in Fig.~\ref{fig:QDsketch}~(c). The letters A, B, C, and D mark the concentrations listed in Tab.~\ref{tab:ABCDdesig}. For the alloy interpolation scheme see Eqs.~(\ref{Eq:alloy_lin-1})~and~(\ref{Eq:alloy_tern-1}). The marks T-I and T-II denote the type of confinement in real space.
\label{fig:QDexcitonFssBd}}
\end{figure}

We now proceed with the fine structure of $X^0$. That is caused in (In,Ga)As/GaAs QDs~\cite{Schliwa:09,Krapek2015,Krapek2016} by the effects of isotropic and anisotropic exchange interaction, which is the case also for the present system. The former causes the energy separation of bright and dark $X^0$ ($\Delta E_{\rm bd}$) while the latter results in FSS of $X^0$.

%
%
The results for our dots are shown in Fig~\ref{fig:QDexcitonFssBd}, again for both GaP and GaAs substrates in left and right panels, respectively. We find FSS of $X^0$ to be in the range of $\sim 180-300$~$\mu$eV and $\Delta E_{\rm bd}$ of $\sim 400-600$~$\mu$eV for both substrates in type-I regime. On the other hand, for type~II those parameters drop to values $\lesssim 100$~$\mu$eV. We note that the calculations of FSS and $\Delta E_{\rm bd}$ shown in Fig.~\ref{fig:QDexcitonFssBd} were performed with two electron and two hole single-particle basis states and expanded the exchange interaction into a multipole series~\cite{Takagahara1993,Krapek2015}.
Following Ref.~\cite{Krapek2015} we considered the following terms of that expansion: monopole-monopole (EX$_0$), monopole-dipole (EX$_1$), and dipole-dipole (EX$_2$). We find that irrespective of the substrate material (GaP or GaAs) the FSS in our system is dominated by EX$_2$. On the other hand, EX$_0$ and EX$_1$ contribute to FSS and $\Delta E_{\rm bd}$ of only $3-10$~$\mu$eV ($<0.5$~$\mu$eV) and $\sim 30$~$\mu$eV ($<1$~$\mu$eV), respectively, for type-I (type-II) confinement. We further note that considerably smaller FSS for type~II corroborates with the results of Refs.~\cite{Krapek2015,Krapek2016,Klenovsky2015} for (In,Ga)As/Ga(As,Sb)/GaAs QDs and, in turn, confirms that to be a rather general property of dots which are type-II in real space.

The correlation is obtained in our CI calculations through admixing of excited single-particle states~\cite{Klenovsky2017}. By taking the basis of two (six) ground state electron and two (six) hole states for calculations without (with) the effects of correlation, we have found the effect on FSS and $\Delta E_{\rm bd}$ energies to be $\sim 2\,\,\mu$eV (not shown).
In total, the above findings make In$_{1-x}$Ga$_x$As$_y$Sb$_{1-y}$ QDs with GaAs IL on GaP substrate a promising candidate for realization of optically bright single photon sources, different to type-I (In,Ga)As/GaAs QDs which are currently being under investigation as sources of light for quantum cryptography applications~\cite{Fischbach2017,Paul2017,Schlehahn2018}.


\begin{figure}[!ht]
\renewcommand{\tabcolsep}{2pt}
\begin{center}
\begin{tabular}{c}
\includegraphics[width=0.48\textwidth]{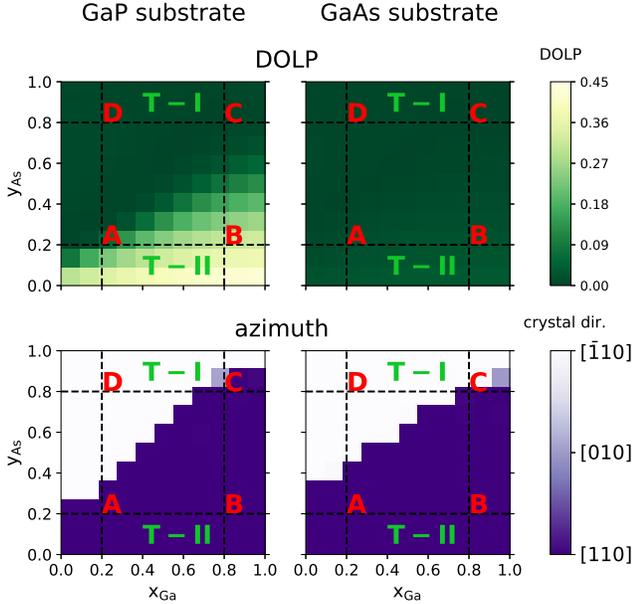}
\end{tabular}
\end{center}
\caption{DOLP and azimuth of the polarization of the emission from the incoherent sum of bright $X^0$ doublet of $\Gamma$ states for In$_{1-x}$Ga$_x$As$_y$Sb$_{1-y}$ QD with GaAs IL grown on GaP (left column) or on GaAs (right column) substrates, respectively. The azimuth angles are shown in terms of the crystal directions. The single-particle basis of CI calculations was two electron and two hole ground states. Except for the composition, QD structural properties were the same as those in Fig.~\ref{fig:QDsketch}~(c). The letters A, B, C, and D mark the concentrations listed in Tab.~\ref{tab:ABCDdesig}. For the alloy interpolation scheme see Eqs.~(\ref{Eq:alloy_lin-1})~and~(\ref{Eq:alloy_tern-1}). The marks T-I and T-II denote the type of confinement in real space. Notice that the azimuth fairly well indicates the type of confinement in our QDs.
\label{fig:QDexcitonDOLP}}
\end{figure}
Furthermore, we would like to provide a useful way of experimental determination the type of confinement in In$_{1-x}$Ga$_x$As$_y$Sb$_{1-y}$/GaAs/GaP QDs based on measurement of the polarization of emission of $X^0$, motivated by Ref.~\cite{Klenovsky2015}. For the incoherent sum of the bright $X^0$ doublet, we show in Fig.~\ref{fig:QDexcitonDOLP} the polarization azimuth and the degree of linear polarization (DOLP), defined by
\begin{equation}
\label{Eq:DOLP}
{\rm DOLP}=\frac{R_{\rm max}-R_{\rm min}}{R_{\rm max}+R_{\rm min}},    
\end{equation}
where $R_{\rm max}$ and $R_{\rm min}$ denote the maximum and minimum value of $R$, respectively. Note that the azimuth is given in terms of the crystallographic axes in order to ease the comparison with the shape of the wavefunctions, shown in Fig.~\ref{fig:QDwfs}. Similarly as in Ref.~\cite{Klenovsky2015}, we find that the azimuth of $X^0$ in type-II regime follows the orientation of the elongation of the wavefunction of the quasiparticle which is outside of the dot body. Contrary to Ref.~\cite{Klenovsky2015}, in the present system the quasiparticles outside of QD are electrons which are elongated along $[110]$ axis, hence, the orientation of the azimuth in type II. In type I, on the other hand, the azimuth is dictated by the anisotropy of hole wavefunctions which is along $[1\overline{1}0]$ axis. Thus, the $90^{\circ}$ flip of the polarization azimuth of emission from In$_{1-x}$Ga$_x$As$_y$Sb$_{1-y}$/GaAs/GaP QDs, when going from type I to type II, is a clear sign of the type of confinement. 

On the other hand, DOLP of the incoherently summed $X^0$ is close to zero in type-I In$_{1-x}$Ga$_x$As$_y$Sb$_{1-y}$ QDs on GaAs IL irrespective of the substrate. However, that is approaching $\sim 0.5$ for type II ($y_{\rm As}\lesssim0.2$) in case of QDs grown on GaP substrate but not on GaAs. This is a consequence of the GaAs IL in In$_{1-x}$Ga$_x$As$_y$Sb$_{1-y}$/GaAs/GaP QDs providing additional confinement for $\Gamma$-electrons which is not present, however, if the substrate is GaAs instead of GaP. 

We note that the values of FSS, $\Delta E_{\rm bd}$ and DOLP, might be slightly different in dots which do not have uniform alloy content or are elongated.

For the sake of completeness, we note that the results corresponding to fine-structure, Fig.~\ref{fig:QDexcitonFssBd}, can be confirmed experimentally,~e.~g., by resonant PL~\cite{Rautert2019,ArxivReindl:19}. The results discussed in Fig.~\ref{fig:QDexcitonDOLP} were in part observed in emission of type-I In$_{1-x}$Ga$_x$As$_y$Sb$_{1-y}$/GaAs/GaP QDs in Ref.~\cite{ArxivSteindl:19}.


\section{Application as quantum gates}

The separation of $\Gamma$-electron wavefunctions that are type-II in real space for structure B (see Tab.~\ref{tab:ABCDdesig})
into two segments, seen in Fig.~\ref{fig:QDwfs}, is qualitatively similar to that occurring for hole states in type-II (In,Ga)As QDs overgrown with Ga(As,Sb) layer~\cite{Klenovsky2010}.
The electron wavefunctions form molecular-like states, in the sense that the four lowest energy states of a complex of two interacting electrons forms a singlet and a triplet~\cite{Burkard1999,Klenovsky2016}. Hence, we tested the proposal of quantum gate (QG) given by Burkard~{\sl et al.} in Ref.~\cite{Burkard1999} on our system. We note that the {\sl qubit} discussed by Burkard~{\sl et al.} is based on the electron {\sl spin} and it works by changing the sign of the exchange energy $J=E_{\rm t}-E_{\rm s}$ \{$E_{\rm t}$ ($E_{\rm s}$) is the energy of triplet (singlet)\} of two electron complex in QD by magnetic flux density ($B$) applied along [001] crystal direction. The necessary requirement for the correct operation of QG under consideration is that the lowest energy state of the two electron complex for $B=0$~T is singlet, i.e., a highly entangled spin state~\cite{Burkard1999}.

We test two In$_{0.5}$Ga$_{0.5}$Sb/GaAs/GaP QD structures: (i) QD1 with properties given in Fig.~\ref{fig:QDsketch}~(c) and (ii) QD2 with base diameter $d_b=15$~nm, height $h=1.5$~nm, and positioned on 3 ML thick GaAs IL. Note that both QD1 and QD2 are defined in GaP substrate and have $y_{\rm As}=0$ and $x_{\rm{Ga}}=0.5$. The choice of $y_{\rm As}$ was made in order to ``push" $\Gamma$-electron wavefunction towards GaAs IL, while $x_{\rm{Ga}}$ is chosen to be some mean content mainly due to the fact that this parameter is not critical for the operation of our QG. We then apply $B$ on QD1 and QD2 in [001] direction, taking into account the Zeeman-Hamiltonian in single-particle eight-band ${\bf k}\cdot{\bf p}$ calculations for $\Gamma$-electrons. Note that due to the multiband ${\bf k}\cdot{\bf p}$ we allowed also for coupling of electrons to $\Gamma$ valence band states. The states of two electron complexes is then computed by CI with four electron single-particle basis states. 

\begin{figure}[!ht]
\renewcommand{\tabcolsep}{2pt}
\begin{center}
\begin{tabular}{c}
\includegraphics[width=0.48\textwidth]{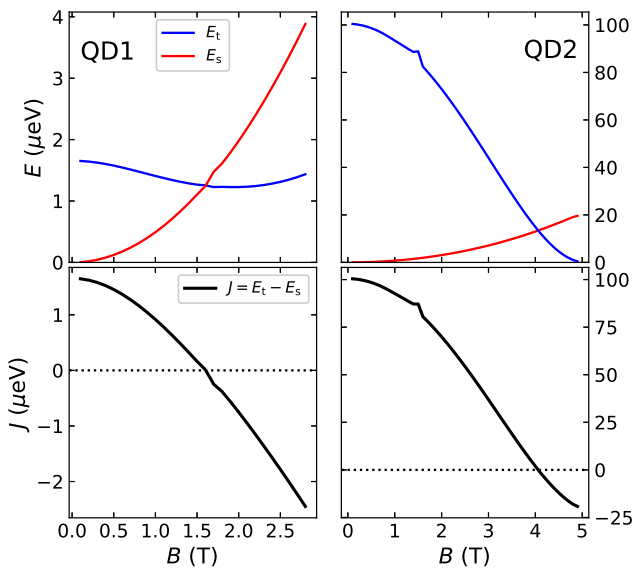}
\end{tabular}
\end{center}
\caption{Upper panels show energies of triplet ($E_{\rm t}$, blue) and singlet ($E_{\rm s}$, red) states of a complex of two interacting electrons while lower panels show the exchange energy $J=E_{\rm t}-E_{\rm s}$. The results in the left two graphs are for QD1 while on the right are those for QD2 (see text for QD1 and QD2 structural parameters). The dotted horizontal line marks zero value of $J$. Note different values of $B$, $E$, and $J$ for QD1 and QD2, respectively.
\label{fig:QG}}
\end{figure}

The results shown in Fig.~\ref{fig:QG} demonstrate that, for both QD1 and QD2, the lower energy state at $B=0$~T is singlet and that one can tune $J$ by increasing $B$ reaching crossing through zero at $B=1.5$~T and $B=4$~T, respectively. Note that, while tuning range of $J$ is considerably larger for QD2, the crossing occurs at larger $B$ as well.

\begin{table}[!ht]
\caption{Comparison of selected parameters between the QD1 and QD2 (see text for their structural parameters) and calculations of Burkard et.~al.~\cite{Burkard1999}. The meaning of the parameters is the following: $J_0$ denotes $J$ for $B=0$~T; $a_B=\sqrt{\hbar/m\omega_{\rm sp}}$ is the effective Bohr radius; $\hbar\omega_{\rm sp}$ is the energy difference between the single-particle electron state belonging to Bloch wave with $s$-symmetry and that with $p$-symmetry; $\tau$ is the ratio of the probability density in the middle between the segments to the peak probability density.\label{tab:MoleculeComparison}}
\begin{tabular}{l|c|c|c}
&QD1&QD2&Ref.~\cite{Burkard1999}\\
\hline
$J_0$ ($\mu{\rm eV}$)& 1.7& 100& 700\\ 
$\hbar\omega_{\rm sp}$ (meV)& 14& 8& 3\\
$a_B$ (nm)& 9.1& 12& 20\\ 
$a/a_B$& 1.2& 0.9& 0.7\\
$\tau$ (\%)& 0.4& 5& $\lesssim$20\\
\end{tabular}
\end{table}

To see the reason for that, we show in table~\ref{tab:MoleculeComparison} the comparison of results for QD1 and QD2. We choose similar parameters as in Ref.~\cite{Klenovsky2016} defined in~\cite{Burkard1999}: $J$ for $B=0$~T denoted by $J_0$; $\hbar\omega_{\rm sp}$ being the energy difference between the single-particle electron state belonging to Bloch wave with $s$-symmetry and that with $p$-symmetry to which the electron might escape,~e.~g.,~due to thermal radiation; the effective Bohr radius $a_B=\sqrt{\hbar/m\omega_{\rm sp}}$ of the two electron complex where $m=0.067m_e$ is the $\Gamma$-point electron effective mass in GaAs~\cite{Vurgaftman2001} and $m_e$ is the mass of free electron; ratio of $a/a_B$ where $a$ is half of the distance between the wavefunction segments; $\tau$ is the ratio of the probability density in the middle between the segments to the maximum probability density, which characterizes the coupling of the electrons. Clearly, QD2 seems to be more favorable for a realization of QG than QD1 which behaves somewhat on the borderline between electron quantum ``molecule" and two uncoupled QDs. We show in Tab.~\ref{tab:MoleculeComparison} also the corresponding values of Burkard~{\sl et al.}~\cite{Burkard1999}. It is interesting to note that $\hbar\omega_{\rm sp}$ roughly corresponds to the maximum operational temperature which can be for QD1 and QD2 obtained by dividing $\hbar\omega_{\rm sp}$ by Boltzmann constant leading to values of $\sim 162$~K and $\sim 92$~K, respectively, both of which are higher than liquid nitrogen temperature.

Due to low coupling of the spins of electrons to that of the atomic nuclei, the In$_{1-x}$Ga$_x$As$_y$Sb$_{1-y}$/GaAs/GaP QD system provides potentially much lower dephasing~\cite{Burkard1999} than the (In,Ga)As/Ga(As,Sb)/GaAs QDs studied in Ref.~\cite{Klenovsky2016}, where QG was based on the spin of holes. However, clear disadvantage of the current proposal lies in the fact that $\Gamma$-electrons are not the ground state for that quasiparticle, see Fig.~\ref{fig:QDexcitonEF} which might influence the way the two electron state is initialized in our QG. One possibility of overcoming that is to put QG into intrinsic part of PIN diode and utilize the effect of quantum tunneling by setting an appropriate voltage similarly as it is done in the QD-Flash memory concept~\cite{t_sala}. Another drawback then, however, lies in the time the two electrons will stay in $\Gamma$-band in IL until they are scattered,~e.~g.,~to ${\bf k}$-indirect states. Here the mixing of those with $\Gamma$ ones will be important and following Ref.~\cite{Diaz2006} that will be unfortunately more pronounced for QD2 because of its smaller size compared to QD1. Nevertheless, we believe that our system is an interesting alternative for QG realization.

%



\section{Conclusions}

Studies of the  electronic structure of In$_{1-x}$Ga$_x$As$_y$Sb$_{1-y}$/GaAs quantum dots grown either on GaP or GaAs substrates are presented. We first determine the confinement potentials for ${\bf k}\neq 0$ and ${\bf k}= 0$ conduction and ${\bf k}= 0$ valence bands. The latter along with the calculated single-particle hole states enable us to determine the most promising candidate structures for the realization of the QD-Flash memory concept from this system. Based on the calculated confinement potentials, we proceed with the determination of single-particle electron and hole states and the energy ordering of their mutual transitions. Here, we thoroughly discuss the method of ${\bf k}\cdot{\bf p}$ calculations for ${\bf k}$-indirect transitions, and determine the form of the momentum matrix element that needs to be determined for such calculations to be correct. For transitions between $\Gamma$-electron and $\Gamma$-hole states we compute also the excitonic states. Through investigation of their emission rates, we identify for which concentrations of dot material constituents type-I or type-II confinement should be expected, and we show FSS and bright-dark splitting including the effect of the multipole expansion of exchange interaction. Moreover, we provide a neat method to experimentally determine the type of confinement from the measurements of the polarization of photoluminescence. Finally, we consider using In$_{1-x}$Ga$_x$As$_y$Sb$_{1-y}$/GaAs/GaP quantum dots as quantum gates and discuss their properties.

In conclusion, comparing to the (In,Ga)As/GaAs system, we show that, despite the presence of ${\bf k}$-indirect transitions, In$_{1-x}$Ga$_x$As$_y$Sb$_{1-y}$/GaAs/GaP quantum dots are perhaps more useful for effective realization of most of the building blocks of quantum information technology based on quantum dots, like entangled-photon sources or qubits. Left for future investigations based on full-zone methods such as the empirical tight-binding are the effects of inter-valley coupling and the calculation of L/X to $\Gamma$ transition probabilities.

\section{Acknowledgements}

P.K. would like to acknowledge the help of Elisa Maddalena Sala, Petr Steindl, and Diana Csontosov\'a with grammar and visual style corrections and for fruitful discussions. A part of the work was carried out under the project CEITEC 2020 (LQ1601) with financial support from the Ministry of Education, Youth and Sports of the Czech Republic under the National Sustainability Programme II. This project has received national funding from the MEYS and the funding from European Union's Horizon 2020 (2014-2020) research and innovation framework programme under grant agreement No 731473. P.K. was supported through the project MOBILITY, jointly funded  by the Ministry of Education, Youth and Sports of the Czech Republic under code 7AMB17AT044. 
The work reported in this paper was (partially) funded by project EMPIR 17FUN06 Siqust. This project has received funding from the EMPIR programme co-financed by the Participating States and from the European Union’s Horizon 2020 research and innovation programme.
%
%

\bibliography{paper_TUB.bib}

\end{document}